\documentclass[10pt,english,aps,prd,nofootinbib,superscriptaddress,preprintnumbers]{revtex4-2}
\usepackage[T1]{fontenc}
\usepackage[latin9]{inputenc}
\usepackage{bm}
\usepackage{amsmath}
\usepackage{amssymb}

\allowdisplaybreaks

\makeatletter
\def\Mpl{M_{\rm P}}

\makeatother

\usepackage{babel}
\begin{document}
\preprint{YITP-21-127, IPMU21-0068}
\title{Static, spherically symmetric objects in Type-II minimally modified gravity}
\author{Antonio De Felice}
\affiliation{Center for Gravitational Physics, Yukawa Institute for Theoretical
Physics, Kyoto University, 606-8502, Kyoto, Japan}
\author{Shinji Mukohyama}
\affiliation{Center for Gravitational Physics, Yukawa Institute for Theoretical
Physics, Kyoto University, 606-8502, Kyoto, Japan}
\affiliation{Kavli Institute for the Physics and Mathematics of the Universe (WPI),
The University of Tokyo, Kashiwa, Chiba 277-8583, Japan}
\author{Masroor C.\ Pookkillath}
\affiliation{Center for Gravitational Physics, Yukawa Institute for Theoretical
Physics, Kyoto University, 606-8502, Kyoto, Japan}
\date{\today}
\begin{abstract}
Static, spherically symmetric solutions representing stars made of barotropic perfect fluid are studied in the context of two theories of type-II minimally modified gravity, VCDM and VCCDM. Both of these theories share the property that no additional degree of freedom is introduced in the gravity sector, and propagate only two gravitational waves besides matter fields, as in General Relativity (GR). We find that, on imposing physical boundary conditions on the Misner-Sharp mass of the system, the solutions in V(C)CDM exactly coincide with the ones in GR, namely they also satisfy the Tolman-Oppenheimer-Volkoff equation. 
\end{abstract}
\maketitle

\section{Introduction}

We live in a particular era in cosmology due to the remarkable precision achieved in various cosmological observations. As a matter of fact, since the discovery of gravitational waves \citep{LIGOScientific:2016aoc}, General Relativity (GR) has been confirmed by yet another independent experimental data. GR then appears more and more to be the theory of classical gravitational interactions. This picture is quite astonishing for a theory that was introduced more than one hundred years ago.

To this beautiful picture and powerful result of theoretical physics, cosmological observations are adding to it several discrepant results. One of the most embarrassing among them is the tension among different measurements of the expansion rate of our universe today, which is called the Hubble parameter $H_{0}$. Although this is one of the oldest and elementary measurements in cosmology, on assuming GR to hold at all times after the big bang, we find that different data sets find different values for this unique observable. The only ways out: experiments are wrong, statistical analysis is not correct, or, surprisingly, the assumed underlining theory does not hold.

If GR is not suspicious in the context of cosmology, then either the cosmological data or the statistical analysis in determining $H_{0}$ cannot be trusted. Evidently, this statement (if correct) needs then to be completed by finding the reason why the data or/and the analysis are wrong. As long as we do not have a clear explanation for it, the other possibilities need to be explored. One could also conclude that the classical theory of gravity as described by GR is correct; what is missing is a proper description of the matter present in the universe. Then, we need to understand better what kind of matter can be responsible for the different values of $H_{0}$ without contradicting with all experiments and observations so far. On top of that, one would like to give possible predictions from the requirement that the extra matter responsible for the tension would be ``visible'' only in cosmology and have no (or negligible) effects otherwise, e.g.\ at solar system scales.

Recently more cosmological data have been adding to this obscure picture another unsolved puzzle \citep{Heymans:2020gsg,DES:2021bvc,Sagredo:2018ahx}. In the context of GR, these data sets tend to give a prediction for the growth rate of the structure which is lower than and in tension with the predictions coming from early-times data sets \citep{Planck:2018vyg}. On top of that, extra matter fields, adding energy to the system, in general, tend to enhance the rate of growth for the matter perturbations and thus to increase the tension. 

The bottom line is that the cosmological picture seems to be not fitting with a simple $\Lambda$CDM model of the universe. Even though we end up with this revolutionary conclusion, still we need then to find a valid alternative to the so-far consensus cosmological model, i.e.\ a new model which is required to overcome the existing puzzles that we are facing in cosmology.

Recently, a new paradigm has been introduced, i.e.\ changing gravity minimally \citep{DeFelice:2015hla}. This implies that we change gravity without introducing any additional degree of freedom on top of what are already present in GR, i.e.\ two tensor modes. This class of modified gravity theories were named {\it minimally modified gravity (MMG)} \citep{Lin:2017oow} and classified into two types \citep{Aoki:2018brq}: type-I theories have the Einstein frame and type-II theories do not. This classification was recently further refined in \citep{Aoki:2021zuy}: type-Ia and type-IIa having the standard dispersion relation $\omega^2=k^2$ for tensorial gravitational waves, and type-Ib and type-IIb having non-standard dispersion relations. While all type-I MMG theories can be systematically constructed by the general method developed in \citep{Aoki:2018zcv}, type-II MMG theories have so far been discovered in a case-by-case manner. The construction of type-II MMG theories can thus be accomplished in several different ways, and in any approaches the number of gravity's degrees of freedom need to be set non-linearly and on a general background. Though not necessary, the construction of these theories typically involves the study of the degrees of freedom in the Hamiltonian formalism of the theory. In several cases, it comes out simpler to introduce theories of MMG directly at the level of the Hamiltonian \citep{Mukohyama:2019unx}. This approach has been implemented following different paths e.g.\ in \citep{DeFelice:2015hla,Lin:2017oow,Aoki:2018zcv,Mukohyama:2019unx}. Further studies were aimed to find whether these models could accommodate cosmological data \citep{Aoki:2018brq,Aoki:2020oqc,DeFelice:2016ufg,DeFelice:2015moy,Bolis:2018vzs,Hagala:2020eax,deAraujo:2021cnd,2021arXiv211001237D}, or to find other models which could be more suitable for cosmology, especially in the light of the puzzles of today's gravity at large scales. In particular, recently a theory of type-II MMG called VCDM was introduced to implement a dynamical component at the level of the cosmological background without introducing any new degree of freedom \citep{DeFelice:2020eju}, which seems to give a promising insight into the $H_{0}$-tension puzzle \citep{DeFelice:2020cpt}. Moreover, for this theory, spherically symmetric vacuum solutions were studied in \citep{DeFelice:2020onz} showing that for compact objects the solutions were the same as in GR. Finally, in \citep{DeFelice:2020prd}, the minimal approach followed in VCDM was extended to have a dark matter component which could have different dynamics from the standard baryonic component. In particular, in this new theory, here and after called VCCDM, such a dark component could feel an effectively different gravity force, a phenomenon which was proposed to explain the weakening of growth of structure as compared to $\Lambda$CDM.

Independent of the above mentioned theories, another theory of MMG named cuscuton has also been introduced in the literature \citep{Afshordi:2006ad} (see also \citep{Iyonaga:2018vnu} for an extended version). This theory can be considered as a infinite speed of sound limit of the k-essence theory. It was established in \citep{Aoki:2021zuy} that, when the divergence of $\lambda_{\rm gf}^i$ vanishes, the prediction of the VCDM theory exactly agrees with that of the cuscuton theory, where $\lambda_{\rm gf}^i$ is the vector-type Lagrange multiplier enforcing the spatial constancy of an auxiliary scalar field. Indeed, if the divergence of $\lambda_{\rm gf}^i$ vanishes (or more generally if the divergence of $\lambda_{\rm gf}^i$ divided by the lapse function is spatially constant) then the equations of motion of the VCDM enforces the constant mean curvature slice $K=K(t)$. This is consistent with the fact that the cuscuton in the unitary gauge is always in the constant mean curvature slice~\footnote{For this reason the cuscuton differs from a low energy limit of the projectable Ho\v{r}ava-Lifshitz gravity \citep{Horava:2009uw}, which possesses three gravitational degrees of freedom, i.e. the tensorial gravitational waves and the ``dark matter as integration constant'' \citep{Mukohyama:2009mz,Mukohyama:2009tp}.}. Therefore the predictions of the two theories agree with each other for a flat FLRW background and linear perturbations around it. On the other hand, if one is to show the equivalence of the two theories, such a configuration, namely $D_i\lambda_{\rm gf}^i=0$, needs to be satisfied globally (i.e.\ over the whole spacetime) and non-linearly, i.e. not only at the level of a homogeneous background and linear perturbations but also at any order in perturbations and even non-perturbatively. Furthermore, the equations of motion of VCDM completely determine the divergence of $\lambda_{\rm gf}^i$ which indeed depends also on the trace of the extrinsic curvature which in turn satisfies its own equations of motion (i.e.\ the modified Einstein equations for the metric components). Therefore, in general the divergence of $\lambda_{\rm gf}^i$ divided by the lapse function depends on the spatial coordinates and this means that the VCDM is not in the constant mean curvature slice. This indeed happens e.g.\ for the general time-dependent background solutions of spherically symmetric vacuum configurations \cite{DeFelice:2020onz}. Therefore it is concluded that the VCDM theory differs from the cuscuton in the unitary gauge. The precise relation between these two theories at the fully nonlinear level remains an open question. 

For VCDM (and VCCDM) spherically symmetric, static and non-static solutions were found in vacuum \citep{DeFelice:2020onz}. In particular, new non-trivial black holes solutions were found which, on the other hand, on setting boundary conditions suitable for compact objects in terms of the Misner-Sharp mass, and matching the effective cosmological constant to the asymptotic cosmological value, are reduced to the standard Schwarzschild de Sitter one of $\Lambda$CDM.

In this paper, we pursue the search for physical solutions of VCDM (and VCCDM) by looking for static and spherically symmetric configurations in the presence of a barotropic baryonic (i.e.\ neglecting dark matter contributions) fluid with some given general equations of state $p=p(\rho)$. In other words, we are looking for static solutions which would show the existence of spherically symmetric stars for these MMG theories. We find indeed that in this case, the solutions for VCDM and VCCDM coincide with each other and are in general different from the ones found in $\Lambda$CDM. However, similarly to the case of black holes, once we impose suitable asymptotic conditions on the Misner-Sharp mass of the system, we re-find, on further neglecting (as in $\Lambda$CDM) the tiny cosmological constant, the standard solutions of the Tolman-Oppenheimer-Volkoff (TOV) equation of GR \citep{Oppenheimer:1939ne,Tolman1987Relativity}. These solutions are, as far as we know, the first exact stellar solutions for the MMG theories. Although the result shows no deviation from GR for static, spherically symmetric configurations satisfying the suitable boundary conditions, still their existence makes these theories suitable to tackle gravity not only at cosmological scales but also at astrophysical scales such as those of compact objects. We hope this work will then motivate further studies into the phenomenology of the MMG theories also in other astrophysical contexts such as rotating compact objects and gravitational collapses.

\section{The Model Description}

In this section we will briefly review the theories of type-II minimally modified gravity (MMG) introduced in \citep{DeFelice:2020eju,DeFelice:2020prd}, the metric ansatz and the description of baryonic matter. 

\subsection{VCDM and VCCDM}

The idea behind the theories is rather simple. The starting point is the Hamiltonian of General Relativity (GR). Then a canonical transformation to another frame is performed via a generating function that depends on new variables and old momenta. At this level, one still has GR in vacuum although written in unconventional variables. At this level, in the new frame either 1) a cosmological constant, or 2) a cosmological constant and a dark matter model of one's choice is/are added. The former gives the VCDM theory and the latter gives the VCCDM theory, after the following steps. On adding 1) or 2) in this frame, GR changes into another theory. On doing this one loses one first-class constraint which reduces to a second-class one. Then we need to introduce a further second-class constraint to keep the same number of degrees of freedom as that of GR in the gravity sector, i.e. to keep the theory minimal. After this step, an inverse canonical transformation is performed. The resulting Hamiltonian is the Hamiltonian of the new theory (either VCDM or VCCDM) written in the original variables introduced at the starting point. Now a Legendre transformation is performed to find the Lagrangian of the new theory. To this Lagrangian, we can safely add other matter Lagrangians including the standard model of particle physics, e.g.\ the radiation and baryon sectors (as well as dark matter in the case of VCDM)\footnote{In principle, one can add all matter fields in the new frame and then, after imposing the extra second class constraint, perform the inverse canonical transformation in order to reach the initial starting frame. This step was done in \citep{Aoki:2018brq}, but it was found the dynamic of this Type-I theory, in the baryonic sector, to be quite strongly constrained.}.

By construction the theory only possesses two tensorial degrees of freedom in the gravity sector together with the matter degrees of freedom. Furthermore, whenever dark matter can be neglected, VCDM and VCCDM share the same phenomenology, i.e.\ the same solutions of the modified Einstein equations. In the following we will discuss solutions of static, spherically symmetric matter fields forming a star. In this case we will consider dark matter to be negligible (as almost pressure-less), and as such VCDM and VCCDM will have the same phenomenology for stars.

Let us now define the Lagrangian of the theory, coupled to a baryonic fluid as the only matter field present in the system. First of all we will make use of the ADM formalism, introducing then the lapse $N$, the shift $N^{i}$ and the three-dimensional metric $\gamma_{ij}$. Out of these variables we can define the extrinsic curvature 
\begin{equation}
K_{ij}\equiv\frac{1}{2N}\,(\dot{\gamma}_{ij}-D_{i}N_{j}-D_{j}N_{i})\,,
\end{equation}
where $D_{i}$ is the three-dimensional covariant derivative compatible with the metric $\gamma_{ij}$. Then we introduce a three-dimensional auxiliary scalar field $\phi$ so that we can build up the following action for the system 
\begin{equation}
S=\int d^{4}x\,N\sqrt{\gamma}\left[\frac{\Mpl^{2}}{2}\,\bigl(R+K_{ij}K^{ij}-K^{2}-2V(\phi)\bigr)-\frac{\lambda_{{\rm gf}}^{i}}{N}\,\Mpl^{2}\,\partial_{i}\phi-\frac{3}{4}\,\Mpl^{2}\lambda^{2}-\Mpl^{2}\lambda\,(K+\phi)+\mathcal{L}_{{\rm baryon}}\right],
\end{equation}
where $R$ is the three-dimensional Ricci scalar of $\gamma_{ij}$, $K\equiv\gamma^{ij}K_{ij}$ is the trace of the extrinsic curvature, $\mathcal{L}_{{\rm baryon}}$ is the matter Lagrangian of the baryonic matter field (which later on, for simplicity, will be modeled by a single perfect fluid component), and $\lambda$ and $\lambda_{{\rm gf}}^{i}$ are instead two Lagrange multiplier fields which are meant to set the constraints which are necessary as to make the theory possess only two tensorial degrees of freedom in the gravity sector. As a consequence of the constraint enforced by $\lambda_{{\rm gf}}^{i}$, $\phi$ is a function of the time only and thus is a global variable. Furthermore, $V(\phi)$ is a free function\footnote{In VCCDM we would have in the dark matter Lagrangian another free function of the scalar $\phi$.}, which then becomes at most a free function of time, and which can be used in order to fulfill some wanted cosmological dynamics \citep{DeFelice:2020eju,DeFelice:2020cpt}.

\subsection{The ansatz}

Let us start with the metric ansatz 
\begin{equation}
\gamma_{ij}dx^idx^j=[F(r)^{2}+\Phi]\,dr^{2}+F_{2}(r)^{2}\,(1+\zeta)\left[\frac{dz^{2}}{1-z^{2}}+(1-z^{2})\,d\varphi^{2}\right],
\end{equation}
where $F$ and $F_{2}$ are general functions of the radial coordinate $r$. Here we have introduced $z\equiv\cos\vartheta$, where $\vartheta$ and $\varphi$ represent the standard polar angles. The two perturbations $\Phi(t,r)$ and $\zeta(t,r)$ are introduced just to find the two background equations of motion, and thus are set to zero after first-order variations of the action with respect to them are computed. 

We introduce the shift vector as 
\begin{equation}
N^{i}\partial_{i}=\left(\frac{B}{F}+\chi\right)\partial_{r}\,,
\end{equation}
where $B=B(r)$ is another free function of $r$ and for VCCDM (or VCDM) needs to be determined by the equations of motion (instead of a gauge condition). We also need to introduce the lapse function 
\begin{equation}
N=N(r)\,(1+\alpha)\,,
\end{equation}
and the three-dimensional auxiliary scalar field 
\begin{equation}
\phi=\phi(r)+\delta\phi\,.
\end{equation}
Similarly to $\Phi$ and $\chi$ introduced above, the variables $\zeta$, $\alpha$ and $\delta\phi$ are perturbations (and functions of $t$ and $r$) which are meant to be used to find the background equations of motion and thus are set to be zero after taking the variations of the action with respect to them. 

Finally we also need to introduce the Lagrange multipliers
\begin{equation}
 \lambda = \lambda(r)+\delta\lambda\,, \quad 
  \lambda^{i}\partial_{i} = \left(\frac{\lambda_{V}(r)}{F}+\delta\lambda_{V}\right)\partial_{r}\,,
 \end{equation}
which are meant to set the constraints for V(C)CDM. Again, the perturbations $\delta\lambda$ and $\delta\lambda_{V}$ are introduced just to find the background equations and thus are set to zero at the end of the calculation. For this static ansatz the extrinsic curvature reduces to 
\begin{equation}
K_{ij}=\frac{1}{2N}\,(D_{i}N_{j}+D_{j}N_{i})\,.
\end{equation}
We are now ready to introduce the matter components in these theories.

\subsection{Baryonic matter fluid}

For the matter here we consider a baryonic matter component, i.e.\ having some nonzero pressure. If we were to consider a pressure-less fluid, assuming CDM to fulfill this assumption, then we would need to study a gravitational collapse, which is time-dependent and inhomogeneous. Although the study of the gravitational collapse is interesting on its own, we will leave it as a future project and instead focus on a different physical issue, namely to find how a star profile changes for these theories with respect to GR. In order to study standard astrophysical objects (neglecting at least for the moment more exotic possibilities), we consider baryonic matter in the form of a single perfect fluid with nonzero pressure, and assume a barotropic equation of state (yet not explicitly fixed) of the kind $p=p(\rho)$. Once we restrict our attention to a standard matter component, i.e.\ a combination of the components already present in the standard model of particle physics, then VCCDM and VCDM solutions coincide, as they differ in the behavior of CDM only. We suppose that the matter fluid respects the spherical symmetry, being then compatible with the symmetries of the three-dimensional metric field, the lapse function, the shift vector, and the remaining fields of V(C)CDM, $\phi$ included, evidently. To this aim we define 
\begin{eqnarray}
J^{0} & = & J^{0}(r)+\delta J\,,\\
J^{i}\partial_{i} & = & [J^{r}(r)+\delta j]\,\partial_{r}\,,
\end{eqnarray}
which together form the four-vector $J^{\mu}.$ Out of these quantities we can build up the four-scalar $J^{\mu}J^{\nu}\,g_{\mu\nu}$, which in the ADM language becomes 
\begin{equation}
J^{\mu}J^{\nu}\,g_{\mu\nu}=-(N^{2}-N_{i}N^{i})\,(J^{0})^{2}+2J^{0}\,(N_{i}J^{i})+J_{i}J^{i}\,,
\end{equation}
where $N_{i}=\gamma_{ij}N^{j},$and $J_{i}=\gamma_{ij}J^{j}$. We can now introduce the number density four-scalar, $n$, as 
\begin{equation}
n=\sqrt{-J^{\mu}J^{\nu}\,g_{\mu\nu}}\,,
\end{equation}
which sets $J^{\mu}$ to be a time-like four-vector. We also need to introduce another scalar, $\ell$, as 
\begin{equation}
\ell=\ell(t,r)+\delta\ell\,,
\end{equation}
where we have also introduced the time dependence as $\ell$ is shift symmetric in the Lagrangian.

We can now introduce the Schutz-Sorkin Lagrangian for the perfect fluid (see e.g.\ \citep{Schutz:1977df,Pookkillath:2019nkn}) as 
\begin{equation}
\mathcal{L}_{{\rm pf}}=-N\sqrt{\gamma}\,\rho(n)-N\sqrt{\gamma}\,(J^{0}\dot{\ell}+J^{i}\partial_{i}\ell)\,,
\end{equation}
which needs to be added to the VCDM (VCCDM) Lagrangian. Since $p=n\,\rho_{,n}-\rho$, then on giving $p=p(\rho$), we also fix $\rho=\rho(n)$, so that we are now ready to find the equations of motion for the system and possibly solve them. Before proceeding further we want to point out here that the four-vector
\begin{equation}
u^{\alpha}=\frac{J^{\alpha}}{n}\,,
\end{equation}
by construction satisfies the constraint $u^{\alpha}u^{\beta}\,g_{\alpha\beta}=-1$, and corresponds to the four-velocity of the fluid element. Then, since we focus on a static solution, staticity is shared also by matter fields so that we need to impose the spatial velocity $u^{i}$ to vanish. In turn this leads to setting 
\begin{equation}
J^{i}\partial_{i}=J^{r}\partial_{r}=0\,,\qquad{\rm or}\qquad J^{r}(r)=0\,.
\end{equation}

\section{Matter field equations in VCDM (VCCDM)}

In the context of VCDM and VCCDM let us consider the spherically symmetric static solutions as produced by a baryonic fluid and how they differ from the standard GR solutions. In GR, it is well known that on considering a static and spherically symmetric perfect fluid with general barotropic equation of state, namely $p=p(\rho)$, where $p$ and $\rho$ are the pressure and the energy density of the fluid respectively, then the system is described by a solution of the Tolman-Oppenheimer-Volkoff (TOV) equations. We need then to find and solve the new equations of motion for the matter fluid in V(C)CDM.

First of all, without loss of generality we can fix the radial coordinate once for all by setting 
\begin{equation}
F_{2}(r)=r\,,
\end{equation}
so that the area of the two-dimensional two sphere, i.e.\ the surface having $r=R={\rm const}.$, becomes $S_{2}=4\pi R^{2}$. Furthermore, since $J^{i}=0$ (in particular $J^{r}=0$), then we have $n=\sqrt{-J^{\mu}J^{\nu}g_{\mu\nu}}=J^{0}\sqrt{-g_{00}}=J^{0}\sqrt{N^{2}-B^{2}}$. We use this condition as to express $J^{0}(r)$ in terms of the number density $n=n(r)$ as 
\[
J^{0}=\frac{n}{\sqrt{N^{2}-B^{2}}}\,,
\]
where we assume that $N^{2}>B^{2}$. Now we can proceed solving the equations of motion for this background. First of all, the equation of motion for $\delta\lambda_{V}$ leads to 
\[
\partial_{r}\phi=0\,,\qquad{\rm or}\qquad\phi=\phi_{0}={\rm constant},
\]
where we have used the assumption of staticity. The equations of motion for $\delta J$ and $\delta j$ lead to 
\begin{eqnarray}
\partial_{t}\ell & = & -\rho_{,n}\sqrt{N^{2}-B^{2}}\,,\\
\sqrt{N^{2}-B^{2}}\,\partial_{r}\ell & = & \rho_{,n}\,B\,F\,,
\end{eqnarray}
respectively. The enthalpy per particle $\mu$ is defined as $\mu=\partial\rho/\partial n$, so we can also write, for instance that $\partial_{t}\ell=-\mu\,\sqrt{N^{2}-B^{2}}$. These equations make sense if and only if $\ell$ satisfies 
\begin{equation}
\ell=-\mu_{0}\,t+\tilde{\ell}(r)\,,
\end{equation}
where $\mu_{0}$ is a constant. In turn this implies that 
\begin{equation}
\frac{\partial\rho}{\partial n}=\mu(r)=\frac{\mu_{0}}{\sqrt{N^{2}-B^{2}}}\,,\label{eq:mat_1}
\end{equation}
and 
\begin{equation}
\tilde{\ell}=\int^{r}\frac{BF\rho_{,n}}{\sqrt{N^{2}-B^{2}}}\,d\bar{r}\,.
\end{equation}
Therefore, so far, these equations exactly coincide with the corresponding ones in GR, when we set a gauge for which $B\neq0$.

The equation of motion for $\delta\lambda$ instead sets the profile of $\lambda(r)$ as 
\begin{equation}
\lambda=-\frac{2}{3}\,\phi_{0}+\frac{2}{3}\,\frac{B'}{NF}+\frac{4}{3}\,\frac{B}{rNF}\,.
\end{equation}
In this case we still have the equations of motion for the fields $\alpha$, $\Phi$, $\delta\phi$, $\zeta$, $\chi$, which we name as $\mathcal{E}_{\alpha}=0$, $\mathcal{E}_{\Phi}=0$ and so on. These equations of motion have been written explicitly in Appendix \ref{sec:Equations-of-motion}. All these equations are not independent. In fact, we can show that 
\begin{equation}
\mathcal{E}_{\zeta}+\frac{r\,[n'(B^{2}-N^{2})+N'nN-n\,B'B]}{2n\,(B^{2}-N^{2})}\mathcal{E}_{\delta J}+\frac{r(F\,B'-B\,F')}{F^{2}}\,\mathcal{E}_{\chi}+\frac{rN'}{2N}\,\mathcal{E}_{\alpha}-rF'\,F\,\mathcal{E}_{\Phi}+\frac{rB}{2F}\left(\mathcal{E}'_{\chi}-\frac{2F^{3}}{B}\,\mathcal{E}'_{\Phi}\right)=0\,,\label{eq:bianchi}
\end{equation}
identically\footnote{Here we have also defined $\mathcal{E}_{\delta J}\equiv\frac{\left(\mu_{0}-\rho_{,n}\sqrt{N^{2}-B^{2}}\right)nr^{2}NF}{\sqrt{N^{2}-B^{2}}}$, which is equivalent to Eq.\ (\ref{eq:mat_1}).}. This last identity shows, for example, that $\zeta$'s equation of motion, $\mathcal{E}_{\zeta}$ is not an independent one, and as such, it can be neglected. Hence, we can consider only the equations of motion $\mathcal{E}_{\alpha}=0$, $\mathcal{E}_{\chi}=0$, and $\mathcal{E}_{\Phi}=0$ coming from the gravity sector. It should be noticed that, the equation of motion $\mathcal{E}_{\delta\phi}=0$, can be used in order to fix the solution for $\lambda_{V}$, as 
\begin{equation}
\lambda_{V}=\frac{1}{3r^{2}}\int_{0}^{r}(3NFV_{,\phi}\bar{r}-2NF\bar{r}\phi_{0}+2B'\bar{r}+4B)\,\bar{r}\,d\bar{r}\,.
\end{equation}
Let us reconsider the equation of motion for the matter fluid written in Eq.\ (\ref{eq:mat_1}). We can use it in order to replace everywhere $\rho_{,n}$ with $\mu(r)$. Furthermore, we also have that $p\equiv n\rho_{,n}-\rho=n\mu-\rho$, so that 
\begin{equation}
dp=\mu\,dn+n\,d\mu-\rho_{,n}dn=n\,d\mu\,,
\end{equation}
which, in turn, leads to the continuity equation 
\begin{equation}
p'=n\,\mu'=n\mu\,\frac{\mu'}{\mu}=(\rho+p)\,\frac{\mu'}{\mu}=(\rho+p)\,[\ln(\mu/\mu_{0})]'=-(\rho+p)\,\frac{NN'-BB'}{N^{2}-B^{2}}\,.\label{eq:contin}
\end{equation}
This is identical, in form, to the expression found also in GR, as expected from the fact that the matter Lagrangian for the baryonic fluid does not get modified in these Type-II MMG theories.

\subsection{TOV equation in VCDM (VCCDM)}

Here we consider the Misner-Sharp mass for the system which we define as 
\begin{equation}
^{4}g^{rr}=\frac{N^{2}-B^{2}}{N^{2}F^{2}}=1-\frac{2G_{N}m}{r}-\frac{\Lambda_{{\rm eff}}}{3}\,r^{2}=1-\frac{m(r)}{4\pi r\Mpl^{2}}-\frac{\Lambda_{{\rm eff}}}{3}\,r^{2}\,,
\end{equation}
where $m(r)$ is the $r$-dependent Misner-Sharp mass and $\Lambda_{{\rm eff}}$ is the effective cosmological constant of the system. The above relation can be used as to perform a field redefinition, e.g.\ replacing the variable $F(r)$ in terms of the new variable $m(r)$ as in 
\begin{equation}
F=\frac{2\sqrt{3}\Mpl}{N}\sqrt{\frac{\pi r\,(N^{2}-B^{2})}{12\pi r\,\Mpl^{2}-3m-4\Lambda_{{\rm eff}}\Mpl^{2}\pi\,r^{3}}}\,.\label{eq:F_vs_m}
\end{equation}
If $B$ were to vanish, then $F$ would become a function of $m$ only. As a next step we consider a linear combination of $\mathcal{E}_{\alpha}+b_{1}\mathcal{E}_{\Phi}+b_{2}\mathcal{E}_{\chi}$, where the coefficients $b_{1,2}$ are chosen as to remove $B''$ and $(B')^{2}$ from it, obtaining the following constraint: 
\begin{equation}
\frac{BB'-NN'}{N^{2}-B^{2}}=\frac{8\pi\,\Lambda_{{\rm eff}}\Mpl^{2}\,r^{3}-12\pi\,r^{3}\,(\rho+p)+3\,(r\,m'-m)}{2r\,(12\pi\,r\Mpl^{2}-4\pi\Lambda_{{\rm eff}}\Mpl^{2}\,r^{3}-3m)}\,.
\end{equation}
On combining this with Eq.\ (\ref{eq:contin}), we obtain the continuity equation written in the following alternative form 
\begin{equation}
p'=(\rho+p)\,\frac{8\pi\,\Lambda_{{\rm eff}}\Mpl^{2}\,r^{3}-12\pi\,r^{3}\,(\rho+p)+3\,(r\,m'-m)}{2r\,(12\pi\,r\Mpl^{2}-4\pi\Lambda_{{\rm eff}}\Mpl^{2}\,r^{3}-3m)}\,.\label{eq:OV_gen}
\end{equation}
This equation can also be solved for $m'$ in terms of $p'$ giving
\begin{equation}
m'=\frac{(24r-8r^{3}\Lambda_{{\rm eff}})\pi\,\Mpl^{2}-6m}{3(\rho+p)}\,p'+\frac{12\pi\,r^{3}(\rho+p)+3m-8\pi\Lambda_{0}\Mpl^{2}\,r^{3}}{3r}\,.\label{eq:m_prime_p_prime}
\end{equation}

Now, let us make the following field redefinition 
\begin{equation}
N=\sqrt{\bar{N}^{2}+B^{2}}\,,\label{eq:N_vs_Nbar}
\end{equation}
so that Eq.\ (\ref{eq:contin}) can be rewritten as 
\begin{equation}
p'=-\frac{(\rho+p)\,\bar{N}'}{\bar{N}}\,,
\end{equation}
which can be solved for $\bar{N}$ giving 
\begin{equation}
\bar{N}=\mathcal{C}_{1}\exp\!\left[-\int^{r}\frac{p'\,{\rm d}r_{1}}{\rho+p}\right].
\end{equation}
In the following calculations we may need to use Eq.\ (\ref{eq:m_prime_p_prime}) for $m'$ as found above. Then, on replacing $p'$ inside $\mathcal{E}_{\alpha}=0$ by the right hand side of Eq.\ (\ref{eq:OV_gen}) we obtain the following equivalent constraint
\begin{equation}
\frac{\mathcal{E}_{\alpha}}{\bar{N}}=(rB'+2\,B)^{2}\left(\Lambda_{{\rm eff}}\Mpl^{2}\pi\,r^{3}-3\pi r\Mpl^{2}+\frac{3m}{4}\right)-9r\left\{ \pi r^{2}\left[\rho+\left(V-\frac{\phi_{0}^{2}}{3}-\Lambda_{{\rm eff}}\right)\Mpl^{2}\right]-\frac{m'}{4}\right\} \mathcal{C}_{1}^{2}{\rm e}^{-\int\!\frac{2p'}{\rho+p}\,{\rm d}r}=0\,.
\end{equation}
The solution to this equation can be written as 
\begin{equation}
B=\frac{\mathcal{C}_{2}}{r^{2}}\pm\frac{3\mathcal{C}_{1}}{r^{2}}\int_{0}^{r}\!\frac{r_{1}^{3/2}\,{\rm d}r_{1}{\rm e}^{-\int^{r_{1}}\!\frac{p'}{\rho+p}\,{\rm d}r_{2}}}{\sqrt{\left(12\,r_{1}-4\,r_{1}^{3}\Lambda_{{\rm eff}}\right)\pi\Mpl^{2}-3m}}\sqrt{m'-4r_{1}^{2}\pi\,\rho-4\left(V-\frac{\phi_{0}^{2}}{3}-\Lambda_{{\rm eff}}\right)\Mpl^{2}\pi\,r_{1}^{2}},
\end{equation}
and to avoid a singularity at the origin, we need to set boundary conditions so that $\mathcal{C}_{2}=0$. Substituting this solution in the $\mathcal{E}_{\chi}=0$ equation of motion as well as using Eq.\ (\ref{eq:m_prime_p_prime}), we obtain the following expression for $m''$: 
\begin{equation}
m''=-\left(\frac{16\pi}{3}\,\Mpl^{2}\,r^{2}\Lambda_{{\rm eff}}+\frac{4m}{r}-16\pi\Mpl^{2}\right)\frac{p'}{\rho+p}+4\pi\,r^{2}\,\rho'-\frac{16\pi\,r\Mpl^{2}\Lambda_{{\rm eff}}}{3}+\frac{2m}{r^{2}}+8\,(\rho+p)\pi\,r\,.
\end{equation}
On replacing $p'$ in this last equation by the expression obtained in Eq.\ (\ref{eq:OV_gen}), we find 
\begin{equation}
m''-\frac{2}{r}\,m'-4\pi r^{2}\,\rho'=0\,.
\end{equation}
This result is indeed considerably simple. In fact, its solution is given by 
\begin{equation}
m=\int^{r}(4\pi\rho+\mathcal{C}_{3})\,r^{2}\,dr+\mathcal{C}_{4}=\int_{0}^{r}4\pi\rho\,r^{2}\,dr+\frac{1}{3}\,\mathcal{C}_{3}\,r^{3}\,,
\end{equation}
where we have set boundary conditions so that $m(r=0)=0$ to avoid a singularity at the origin. Notice that the constant $\mathcal{C}_{3}$ induces a new contribution to the Misner-Sharp mass. However, for a compact object it is unphysical to let $m$ grow up to infinity as the distance to the compact object source increases. In fact, we would rather like to set $m$ to stop growing after $p$ (or, equivalently, $\rho$ by virtue of the equation of state) vanishes. On using this extra boundary condition, we are then bound to set $\mathcal{C}_{3}=0$. On doing this last step, we finally obtain the same results as in GR (in the presence of a nonzero cosmological constant and matter).

Then the final TOV equation, a direct consequence of Eq.\ (\ref{eq:OV_gen}), can be rewritten as 
\begin{equation}
p'=-(\rho+p)\,\frac{3\,r^{3}\,p+3\int_{0}^{r}\rho\,r_1^{2}\,dr_1-2\,\Lambda_{{\rm eff}}\Mpl^{2}\,r^{3}}{2r\,(3\,r\Mpl^{2}-\Lambda_{{\rm eff}}\Mpl^{2}\,r^{3}-3\int_{0}^{r}\rho\,r_2^{2}\,dr_2)}\,,\label{eq:TOV}
\end{equation}
which exactly matches the GR result. In fact, we can also immediately write down that 
\begin{equation}
\bar{N}=\mathcal{C}_{1}\exp\!\left[\int_0^{r}\frac{3\,r_{1}^{3}\,p+3\int_{0}^{r_{1}}\rho\,r_{2}^{2}\,{\rm d}r_{2}-2\,\Lambda_{{\rm eff}}\Mpl^{2}\,r_{1}^{3}}{6\,r_{1}\Mpl^{2}-2\Lambda_{{\rm eff}}\Mpl^{2}\,r_{1}^{3}-6\int_{0}^{r_{1}}\rho\,r_{2}^{2}\,{\rm d}r_{2}}\,\frac{{\rm d}r_{1}}{r_{1}}\right],\label{eq:Nbar_sol}
\end{equation}
which again matches the GR results (see Appendix \ref{sec:GR-case}). As for the variable $B$, its general expression for a compact object can be rewritten as 
\begin{equation}
B=\pm\frac{6\mathcal{C}_{1}\Mpl}{r^{2}}\sqrt{\left(\Lambda_{{\rm eff}}+\frac{\phi_{0}^{2}}{3}-V\right)\pi}\int_{0}^{r}\!\frac{r_{1}^{5/2}\,{\rm d}r_{1}{\rm e}^{-\int_0^{r_{1}}\!\frac{p'}{\rho+p}\,{\rm d}r_{2}}}{\sqrt{\left(3\,r_{1}-r_1^{3}\Lambda_{{\rm eff}}\right)4\pi\Mpl^{2}-3m}},
\end{equation}
where $p'$ needs to be replaced by the right hand side of the TOV equation, Eq.\ (\ref{eq:TOV}).

As for the value of the effective cosmological constant $\Lambda_{{\rm eff}}$, being a constant, its value needs to remain unchanged well outside the compact object. This implies that its value should correspond to the effective cosmological constant found for the general static vacuum exterior solution given in \citep{DeFelice:2020onz}, namely 
\begin{equation}
\Lambda_{{\rm eff}}=\Lambda_{{\rm eff},{\rm exterior}}\equiv\Lambda_{0}+3b_{0}^{2}=V(\phi_{0})-\frac{\phi_{0}^{2}}{3}+3b_{0}^{2}\,,
\end{equation}
where $b_{0}=-\frac{1}{3}\,K_{{\rm exterior}}$ is proportional
to the trace of the extrinsic curvature tensor for the exterior vacuum
static solution. Instead $\Lambda_{0}$ corresponds to the value of
the cosmological constant on the homogeneous and isotropic background,
see e.g.\ \citep{DeFelice:2020eju}. In this case we can rewrite
$B$ as follows 
\begin{equation}
B=\pm\frac{9\mathcal{C}_{1}b_{0}\Mpl}{r^{2}}\int_{0}^{r}\!\frac{r_{1}^{5/2}\,{\rm d}r_{1}{\rm e}^{-\int_0^{r_{1}}\!\frac{p'}{\rho+p}\,{\rm d}r_{2}}}{\sqrt{\left[9\,r_{1}-\,r_1^{3}\left(3V(\phi_{0})-\phi_{0}^{2}+9b_{0}^{2}\right)\right]\Mpl^{2}-9\int_{0}^{r_{1}}\rho\,r_{3}^{2}\,dr_{3}}}.\label{eq:B_sol}
\end{equation}
Furthermore, on setting $b_{0}=0$ as to make $\Lambda_{{\rm eff}}=\Lambda_{0}$, then we obtain $B(r)=0$ anywhere inside the star. However, this would in general also set $B$ to be vanishing for the exterior solution. In any case, the formal expression of the TOV does not change, except for replacing the term $\Lambda_{{\rm eff}}$ with $\Lambda_{0}$.

\subsection{Regularity at the origin}

Here we show that at the origin, $r=0$, the solutions are regular. Let us expand the matter density inside the star around the origin as 
\begin{equation}
\rho=\rho_{0}+\rho_{1}r+\frac{1}{2}\rho_{2}\,r^2+\frac{1}{6}\rho_{3}r^{3}+\mathcal{O}\left(r^{4}\right)\,.\label{eq:rho_at_r0}
\end{equation}
Considering an equation of state for the fluid, we can write $p=p(\rho)$. Then expanding the TOV equation Eq.\ (\ref{eq:TOV}) at the origin, comparing coefficients order by order we can find that 
\begin{align}
\rho_{1} & =\rho_{3}=0\,,\\
\rho_{2} & =\left(\rho_{0}+p_{0}\right)\frac{2\Lambda_{\text{eff}}\Mpl^{2}-3p_{0}-\rho_{0}}{6\Mpl^{2}\,p_{,\rho}(\rho_{0})}\,.
\end{align}
This allow us to write the pressure expanded at the origin in the form
\begin{equation}
p=p_{0}+\frac{1}{2}\,p_{,\rho}(\rho_{0})\,\rho_{2}\,r^{2}+\mathcal{O}\left(r^{4}\right)\,.\label{eq:p_at_r0}
\end{equation}

Now, using Eq.\ (\ref{eq:rho_at_r0}) and Eq.\ (\ref{eq:p_at_r0}) in the solution of $\bar{N}$ and $B$, i.e. Eq.\ (\ref{eq:Nbar_sol}) and Eq.\ (\ref{eq:B_sol}) respectively, we find that 
\begin{align}
\bar{N} & =\mathcal{C}_{1}\left[1-\frac{2\Mpl^{2}\Lambda_{\text{eff}}-3p_{0}-\rho_{0}}{12\Mpl^{2}}r^{2}+\mathcal{O}\left(r^{4}\right)\right]\,,\\
B & =\pm\mathcal{C}_{1}\sqrt{3\Lambda_{\text{eff}}+\phi_{0}^{2}-3V}\left[\frac{r}{3}+\frac{1}{20}\frac{\rho_{0}+p_{0}}{\Mpl^{2}}\,r^{3}+\mathcal{O}\left(r^{5}\right)\right]\,.
\end{align}
Finally we can show that 
\begin{equation}
F=1-\frac{\Mpl^{2}\left(\phi_{0}^{2}-3V\right)-3\rho_{0}}{18\Mpl^{2}}\,r^{2}+\mathcal{O}\left(r^{4}\right)\,,
\end{equation}
where we have used Eq.\ (\ref{eq:F_vs_m}) and Eq.\ (\ref{eq:N_vs_Nbar}). Thus, $\bar{N}$ and $F$ are even in $r$ (and $F=1$ in the origin) while $B$ is odd. This shows that the solutions are regular at the origin. 

\section{Conclusions}

In these last years, the phenomenology of gravity has become very interesting in theoretical physics. Evidently, in the context of cosmology, the possibility that $\Lambda$CDM would not be able to explain all the most recent data --- not only the value of today's Hubble parameter, but also the lower growth rate in galaxy surveys --- has forced us to look either into unknown systematics or into new theories which are meant to change the dark sector, i.e.\ to replace the cosmological constant with some as yet unknown new component. Furthermore, the discovery of gravitational waves has paved the way to a plethora of possibilities that are instead oriented to constraining gravity theories on astrophysical scales, i.e.\ very different scales from the cosmological ones. The signals to probe gravity at small scales we have received so far are not pointing to anything different from General Relativity (GR). This implies that, even though gravity does not behave as $\Lambda$CDM at large, cosmological scales, nonetheless, at astrophysical scales, any modified Einstein-Hilbert action is required to be giving astrophysical solutions which match, within our experimental sensitivities, the ones found in GR. On the other hand, one would, in general, expect that either adding new matter fields or modifying the gravity theory, would lead to some modifications at various scales. Therefore several studies have been proposed to make modified gravity theories or extra matter fields not changing the typical GR solutions at the solar system scales. This has implied either the introduction of a very light scalar field (e.g.\ quintessence) almost decoupled from the standard matter sector or the constructions of mechanisms able to screen any extra scalar force (chameleon and Vainshtein mechanisms, for instance).

On the other hand, more recently, \citep{Aoki:2018brq,Mukohyama:2019unx,DeFelice:2020eju,DeFelice:2020prd}, another approach has been put forward in modified theories of gravity. The idea for all these models is to modify gravity without introducing any additional gravitational degree of freedom. This excludes quintessence, vector theories or dRGT massive gravity \citep{deRham:2010ik,deRham:2010kj}, for instance. The new class of theories, on the one hand, should have the same (two tensorial) degrees of freedom as in GR, but on the other hand, it should be different from GR: namely, it should have a different (and possibly interesting) gravitational phenomenology (see e.g.\ \citep{DeFelice:2016ufg,Hagala:2020eax,Aoki:2020oqc,DeFelice:2020cpt,deAraujo:2021cnd}). In particular, for the two theories (VCDM \citep{DeFelice:2020eju} and VCCDM \citep{DeFelice:2020prd}), it was shown recently that in the absence of matter fields, both time-dependent and static spherically symmetric black holes solutions reduce to the standard Schwarzschild de Sitter solutions of GR, provided that physical boundary conditions on the Misner-Sharp mass describing a compact source are imposed \citep{DeFelice:2020onz}.

In this paper, we have explored the existence and properties of static spherically symmetric solutions sourced by a barotropic perfect fluid with an equation of state $p=p(\rho)$, for both VCDM and VCCDM. In other words, we have looked for spherically symmetric stars for these theories of type-II MMG. We also found that, once we set the Misner-Sharp mass to satisfy physical boundary conditions suitable for a compact object (namely it should stop increasing outwards outside the matter source, the stars for these theories reduce to the ones in GR, namely those described by the Tolman-Oppenheimer-Volkoff equation.

As mentioned in the introduction, the relation between the VCDM theory and the cuscuton theory at the nonlinear level is an open question. Considering the fact that time-dependent nonlinear solutions in the VCDM are not in general in the constant mean curvature slice and thus are not solutions of the cuscuton in the unitary gauge, it is important to study the corresponding solutions in the cuscuton theories by extending the analysis in \citep{Iyonaga:2021yfv}. 

This result on one side sheds light on the general properties of these theories and how they differ (or not) from GR. On the other side this result should push forward the search for new properties, i.e.\ new phenomenology, for astrophysical objects/processes in V(C)CDM.

\begin{acknowledgments}
The work of A.D.F.\ was supported by Japan Society for the Promotion of Science Grants-in-Aid for Scientific Research No.\ 20K03969. S.M.'s work was supported in part by Japan Society for the Promotion of Science Grants-in-Aid for Scientific Research No.\ 17H02890, No.\ 17H06359, and by World Premier International Research Center Initiative, MEXT, Japan. The work of M.C.P.\ was supported by the Japan Society for the Promotion of Science Grant-in-Aid for Scientific Research No.\ 17H06359, and also acknowledges the support from the Japanese Government (MEXT) scholarship for Research Student during the initial phase of this project. 
\end{acknowledgments}

\appendix

\section{Equations of motion\label{sec:Equations-of-motion}}

We report here the equations of motion: 
\begin{eqnarray}
\mathcal{E}_{\alpha} & \equiv & \frac{1}{3\left(N^{2}-B^{2}\right)^{\frac{3}{2}}N\,F^{2}}\,\bigl\{\bigl[3r^{2}F^{3}N^{2}(B^{2}-N^{2})\rho-3\,F^{3}N^{4}r^{2}n\rho_{,n}+\Mpl^2(B^{2}-N^{2})\{[3N^{2}+(B'r-B)^{2}]F\nonumber \\
 & - & N^{2}\left(\phi_{0}^{2}r^{2}-3r^{2}V+3\right)F^{3}-6F'N^{2}r\}\bigr]\sqrt{N^{2}-B^{2}}-3r^{2}\,F^{3}N^{2}\mu_{0}n(B^{2}-N^{2})\bigr\}=0\,,\\
\mathcal{E}_{\delta J} & \equiv & \frac{\left(\mu_{0}-\rho_{n}\sqrt{N^{2}-B^{2}}\right)r^{2}nNF}{\sqrt{N^{2}-B^{2}}}=0\,,\\
\mathcal{E}_{\chi} & \equiv & \frac{2\Mpl^{2}(N^{2}-B^{2})\bigl[\{(rF'-2F)B+r[(rB''+2B')F-rB'F']\}N-rFN'(B'r-B)\bigr]-3F^{3}B\,N^{3}r^{2}n\rho_{n}}{3F\,N^{2}\,(B^{2}-N^{2})}=0\,,\\
\mathcal{E}_{\Phi} & \equiv & \frac{1}{6\left(N^{2}-B^{2}\right)^{\frac{3}{2}}N^{2}F^{4}}\Bigl(-3F^{3}\mu_{0}n\left(B^{2}-N^{2}\right)N^{3}r^{2}+\bigl[3F^{3}N^{3}(B^{2}-N^{2})r^{2}\rho+3F^{3}\,B^{2}N^{3}r^{2}n\rho_{,n}\nonumber \\
 & + & \bigl\{[(3r^{2}V-r^{2}\phi_{0}^{2}-3)F^{3}+3F]N^{3}+6rN'F\,N^{2}+\{[(2r^{2}B''+2B'r)B+r^{2}(B')^{2}-3B^{2}]F-2rF'B(B'r-B)\}N\nonumber \\
 & - & 2rN'BF(B'r-B)\bigr\}(B^{2}-N^{2})\Mpl^{2}\bigr]\sqrt{N^{2}-B^{2}}\Bigr)=0\,,\\
\mathcal{E}_{\delta\phi} & \equiv & -\Mpl^{2}r\left[rN\left(V_{\phi}-\frac{2\phi_{0}}{3}\right)F+\frac{2B'r}{3}-r\lambda'_{\mathit{V}}+\frac{4B}{3}-2\lambda_{V}\right]=0\,,\\
\mathcal{E}_{\zeta} & \equiv & \frac{2r}{3\sqrt{N^{2}-B^{2}}\,F^{2}N^{2}}\left\{ \frac{3F^{3}\mu_{0}nN^{3}r}{2}+\sqrt{N^{2}-B^{2}}\left[-\frac{3F^{3}N^{3}\rho r}{2}\right.\right.\nonumber \\
 & + & \Mpl^{2}\left(\left[\frac{r(\phi_{0}^{2}-3V)F^{3}}{2}+\frac{3F'}{2}\right]N^{3}+\left[\frac{3N'F'r}{2}-\left(\frac{3rN''}{2}+\frac{3N'}{2}\right)F\right]N^{2}\right.\nonumber \\
 & + & \left.\left.\left.\left\{ \left[\left(\frac{rB''}{2}-B'\right)B+(B')^{2}r\right]F-\frac{F'B(B'r-B)}{2}\right\} N-\frac{N'BF(B'r-B)}{2}\right)\right]\right\}\,.
\end{eqnarray}
As shown in Eq.\ (\ref{eq:bianchi}), the $\mathcal{E}_{\zeta}$
equation is not independent of the others.

\section{GR case \label{sec:GR-case}}

In GR, for a static spherically symmetric ansatz, we can always make a coordinate transformation to new coordinates $(T, R)$ so that the component $g_{TR}=0$. So, in the following, we consider the coordinates $(T,R)$ as those in which the metric is diagonal and suppose that they are related to the coordinates $(t,r)$ in which the metric is non-diagonal by the following coordinate transformation 
\begin{align*}
t & =T+V(R)\,,\\
r & =R\,.
\end{align*}
In this case we find that the corresponding 1-forms are related to each other by 
\begin{eqnarray}
\bm{d}t & = & \bm{d}T+V_{,R}\,\bm{d}R\,,\\
\bm{d}r & = & \bm{d}R\,.
\end{eqnarray}
On the other hand, we have the following relation between the vector basis,
\begin{eqnarray}
\frac{\partial}{\partial t} & = & \frac{\partial}{\partial T}\,,\\
\frac{\partial}{\partial r} & = & \frac{\partial}{\partial R}-V_{,R}\,\frac{\partial}{\partial T}\,.
\end{eqnarray}
Let us now consider a static matter source. In this case we find 
\begin{equation}
\bm{u}=u^{\alpha}\,\bm{e}_{\alpha}=u^{0}(R)\,\frac{\partial}{\partial T}=u^{0}(r)\,\frac{\partial}{\partial t}\,,
\end{equation}
so that imposing a static matter, $u^{R}=0$, leads to $u^{r}=0$, even after we change the vector basis. On the other hand $u^{0}$ is found via the scalar constraint (which can be evaluated in any coordinate frame): $\bm{g}(\bm{u},\bm{u})=-1$.

Then, since in the $(t,r)$ coordinates the metric is non diagonal, let us consider the 4D metric written in the following form 
\begin{equation}
ds_{4}^{2}=-(N^{2}-B^{2})\,dt^{2}+2BF\,dt\,dr+F^{2}\,dr^{2}+r^{2}\left[\frac{dz^{2}}{1-z^{2}}+(1-z^{2})\,d\tilde{\varphi}^{2}\right].
\end{equation}
Then in GR, we find that the equation of motion for $\delta g_{tr}$ is equivalent to the one for $\delta g_{rr}$. On the other hand, we can also follow the route chosen in the main text for VCDM (or VCCDM), and define $F$ in terms of the Misner-Sharp mass $m$, finding the same result as in Eq.\ (\ref{eq:F_vs_m}). Then the Einstein equations lead to 
\begin{equation}
m'=4\pi r^{2}\rho\,,
\end{equation}
as expected. Furthermore, on defining $\bar{N}=\sqrt{N^{2}-B^{2}}$ as we have also done in the main text, we find that $B$ does not appear any longer in the equations of motion (i.e.\ its dynamics decouples, as expected from the fact that we have a gauge choice for this degree of freedom), and we finally find 
\begin{equation}
\bar{N}=\bar{\mathcal{C}_{0}}\exp\!\left[\int^{r}\frac{2\,\Lambda\,\Mpl^{2}\,r_{1}^{3}-3\,r_{1}^{3}\,p-3\int^{r_{1}}\!r_{2}^{2}\rho\,{\rm d}r_{2}}{2\,\Lambda\,\Mpl^{2}\,r_{1}^{3}-6\,r_{1}\Mpl^{2}+6\int^{r_{1}}\,r_{2}^{2}\rho\,\,{\rm d}r_{2}}\frac{dr_{1}}{r_{1}}\right],
\end{equation}
together with 
\begin{equation}
p'=-\frac{(\rho+p)\,(2\,r^{3}\Lambda\,\Mpl^{2}-3\,pr^{3}-3\,\int\!r^{2}\rho\,{\rm d}r)}{2\,r\,[3\,\int\!r^{2}\rho\,{\rm d}r+r\Mpl^{2}(\Lambda\,r^{2}-3)]}\,.
\end{equation}
So in GR the variable $B$ can be set to any value we like, but the solution for $p,\rho$, and $\bar{N}$ will not change in any way. Instead, in V(C)CDM, $B$ is deduced by solving the equations of motion.

\bibliographystyle{apsrev}
\bibliography{stars}

\end{document}